\documentclass[aps,twocolumn,secnumarabic,balancelastpage,amsmath,amssymb,nofootinbib,hyperref=pdftex]{revtex4}

\usepackage{chapterbib}    % allows a bibliography for each chapter (each labguide has it's own)
\usepackage{color}         % produces boxes or entire pages with colored backgrounds
\usepackage{graphics}      % standard graphics specifications
\usepackage[pdftex]{graphicx}      % alternative graphics specifications
\usepackage{longtable}     % helps with long table options
\usepackage{epsf}          % old package handles encapsulated post script issues
\usepackage{bm}  % special 'bold-math' package
\usepackage{float}
\usepackage{verbatim}
\usepackage{svg}
\usepackage{soul}          % for comment environment
\usepackage[colorlinks=true]{hyperref}  % this package should be added after all others
\usepackage{amssymb}

\UseRawInputEncoding

\begin{document}
\title{Dislocation and Indium Droplet Related Emission Inhomogeneities in InGaN LEDs}
\author {Len van Deurzen$^{1,}$}
\email{lhv9@cornell.edu}
\author{Mikel G\'omez Ruiz$^{2}$}
\author{Kevin Lee$^{3}$}
\author{Henryk Turski$^{4}$}
\author{Shyam Bharadwaj$^{3}$}
\author{Ryan Page$^{5}$}
\author{Vladimir Protasenko$^{3}$}
\author{Huili (Grace) Xing$^{3,5}$}
\author{Jonas L\"ahnemann$^{2}$}
\author{Debdeep Jena$^{1,3,5}$}
%\date{\today}
\affiliation{$^{1}$Department of Applied and Engineering Physics, Cornell University, Ithaca, USA}
\affiliation{$^{2}$Paul-Drude-Institut f\"ur Festk\"orperelektronik, Leibniz-Institut im Forschungsverbund Berlin e.V., Germany}
\affiliation{$^{3}$Department of Electrical and Computer Engineering, Cornell University, Ithaca, USA}
\affiliation{$^{4}$Institute of High Pressure Physics, Polish Academy of Sciences, Warsaw, Poland}
\affiliation{$^{5}$Department of Materials Science and Engineering, Cornell University, Ithaca, USA}

\begin{abstract}
This report classifies emission inhomogeneities that manifest in InGaN quantum well blue light-emitting diodes grown by plasma-assisted molecular beam epitaxy on free-standing GaN substrates. By a combination of spatially resolved electroluminescence and cathodoluminescence measurements, atomic force microscopy, scanning electron microscopy and hot wet KOH etching, the identified inhomogeneities are found to fall in four categories. Labeled here as type~I through IV, they are distinguishable by their size, density, energy, intensity, radiative and electronic characteristics and chemical etch pits which correlates them with dislocations. Type~I exhibits a blueshift of about 120~meV for the InGaN quantum well emission attributed to a perturbation of the active region, which is related to indium droplets that form on the surface in the metal-rich InGaN growth condition. Specifically, we attribute the blueshift to a decreased growth rate of and indium incorporation in the InGaN quantum wells underneath the droplet which is postulated to be the result of reduced incorporated N species due to increased N$_{2}$ formation. The location of droplets are correlated with mixed type dislocations for type I defects. Types II through IV are due to screw dislocations, edge dislocations, and dislocation bunching, respectively, and form dark spots due to leakage current and nonradiative recombination.
\end{abstract}

\maketitle

%%%%%%%%%%%%%%%%%%%%%%%%%%%%%%%%%%%%%%%%%%%%%%%%%%%%%%%%%%%%%%%%%%

The development of free-standing GaN substrates in the last decade has enabled significant progress in GaN-based radio frequency and power device technology \cite{sun_high-performance_2017,kruszewski_algangan_2014,cao_high-power_2004,disney_vertical_2013}, as well as enhanced efficiencies in light-emitting diodes (LEDs) and laser diodes (LDs) \cite{skierbiszewski_high_2005,cich_bulk_2012,bharadwaj_enhanced_2020,lee_light-emitting_2020}. Contributing to these improvements is the reduced number of threading dislocations in homoepitaxial heterostructures, down millionfold to $\approx 10^{3}$~cm$^{-2}$ on bulk substrates from $\gtrsim 10^{8}$~cm$^{-2}$ on sapphire or silicon \cite{mikawa_recent_2020}. Indeed, threading dislocations alter the electronic and radiative properties of the semiconductor and act as scattering centers for mobile carriers or optical modes in photonic devices and waveguides. During a long and turbulent history of study, it has been proposed that edge, screw and mixed type dislocations in wurtzite InGaN can form radiative or nonradiative recombination centers \cite{miyajima_threading_2001,cherns_edge_2001,albrecht_nonradiative_2008,hino_characterization_2000,liu_exciton_2016,elsner_theory_1997,sugahara_direct_1998,lahnemann_carrier_2020}. On the other hand, for indium-containing InAlGaN LEDs, it has been suggested that dislocations do not affect the internal quantum efficiency (IQE) of spontaneous emitters greatly due to the short exciton lifetime and small carrier diffusion length, and due to the lateral confinement induced by compositional fluctuations \cite{chichibu_origin_2006, mishra_unlocking_2021}. Moreover, it is known that mixed and screw dislocations can act as leakage paths in GaN p-n diodes and LEDs \cite{usami_correlation_2018-1, wang_all_2020, li_dependence_2004}. Recently, Usamie et al. and Nakano et al. have shown that Mg diffuses along the dislocation thread and forms an n-type magnesium-threading screw dislocation (Mg-TSD) complex, turning a p-n diode into a n-n conductive leakage channel \cite{usami_direct_2019,nakano_screw_2020}. With the extensive studies that have been reported in the last twenty years, an interesting and still open question is why the IQE of spontaneous emitters grown by molecular beam epitaxy (MBE) is relatively low when compared to the IQE of emitters grown by metal organic chemical vapor deposition (MOCVD).

%====================================================================
\begin{figure*}[t]
\includegraphics[width=16cm]{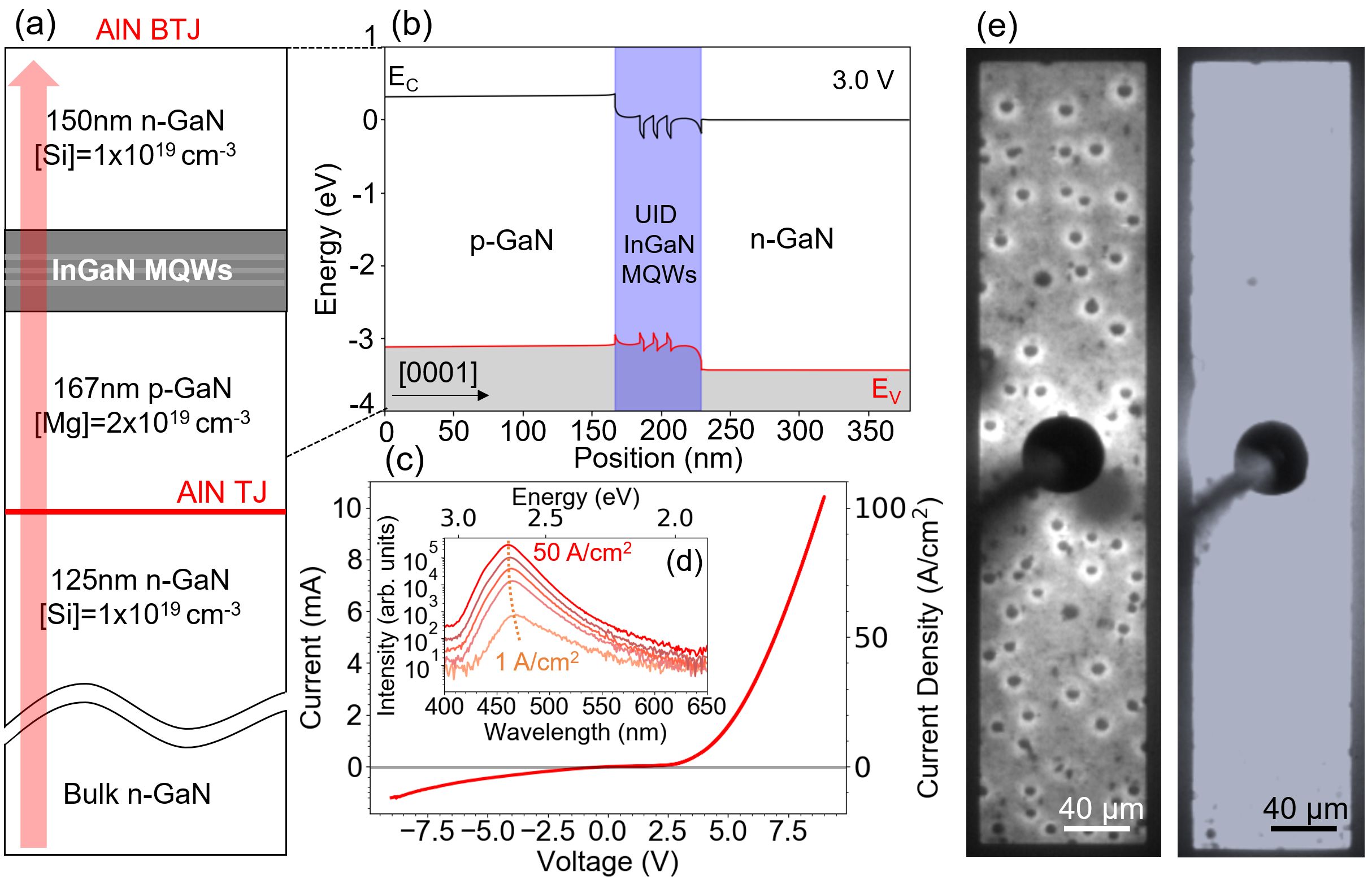}
\caption{AlN-interlayered bottom tunnel junction (BTJ) LED and its device characteristics. (a) Sketch of the quantum heterostructure, (b) energy-band diagram of the p-i-n LED segment at 3.0~V forward bias, (c) the I-V and J-V characteristics of a $100\times100~\mu$m$^{2}$ device, (d) spatially-integrated electroluminescence spectrum at room temperature and various current densities, and (e) optical microscope images of a $100\times500~\mu$m$^{2}$ device in its electroluminescent state at 20~A/cm$^{2}$ (left) and 50~A/cm$^{2}$ (right). \label{fig:Figure_1}}
\end{figure*}
%====================================================================

In this letter, we identify the radiative properties of inhomogeneities in fully-processed blue InGaN LEDs grown by plasma-assisted molecular beam epitaxy (PA-MBE) that result from $a$-, $c$- and ($a$+$c$)-axis dislocations by performing complementary electroluminescence and cathodoluminescence studies. Specifically, the localized emission spectra reveal the electronic and radiative nature of the individual type of defect. We first identify blueshifted emission inhomogeneities which are due to changes to the active region stemming from altered growth conditions underneath indium droplets that form during metal-rich, PA-MBE InGaN growth. Their locations are often correlated with ($a$+$c$)-axis (mixed type) dislocations. We then show that $c$-axis (screw) dislocations are optically dark when relying on electrical injection (by electroluminescence), but not necessarily when relying on local electron-hole pair generation (by cathodoluminescence), which are in essence distinct transport phenomena.  Then, we empirically confirm that the more numerous $a$-axis (edge) dislocations in GaN and InGaN are optically dark, proving their theoretically predicted nonradiative nature. Finally, we discuss regions of dislocation bunching that form V-pit clusters after hot KOH etching and which also exhibit a discrepancy between the electroluminescent and cathodoluminescent state, indicating an alteration of electronic characteristics of the LED cladding layers.  While this work does not pinpoint the reason for the lower efficiencies of MBE LEDs, it advances the quest towards that understanding by identifying the roles of many defects quantitatively. The findings of the emission inhomogeneities in this report, including monochromatic maps, their measured size and density, cathodoluminescence and electroluminescence characteristics, and corresponding explanations are summarized in Table \ref{Table:1}.  

%====================================================================
\begin{table}[t]
\caption{Summary of the dislocation-correlated emission inhomogeneities.
\includegraphics[width=8.7cm]{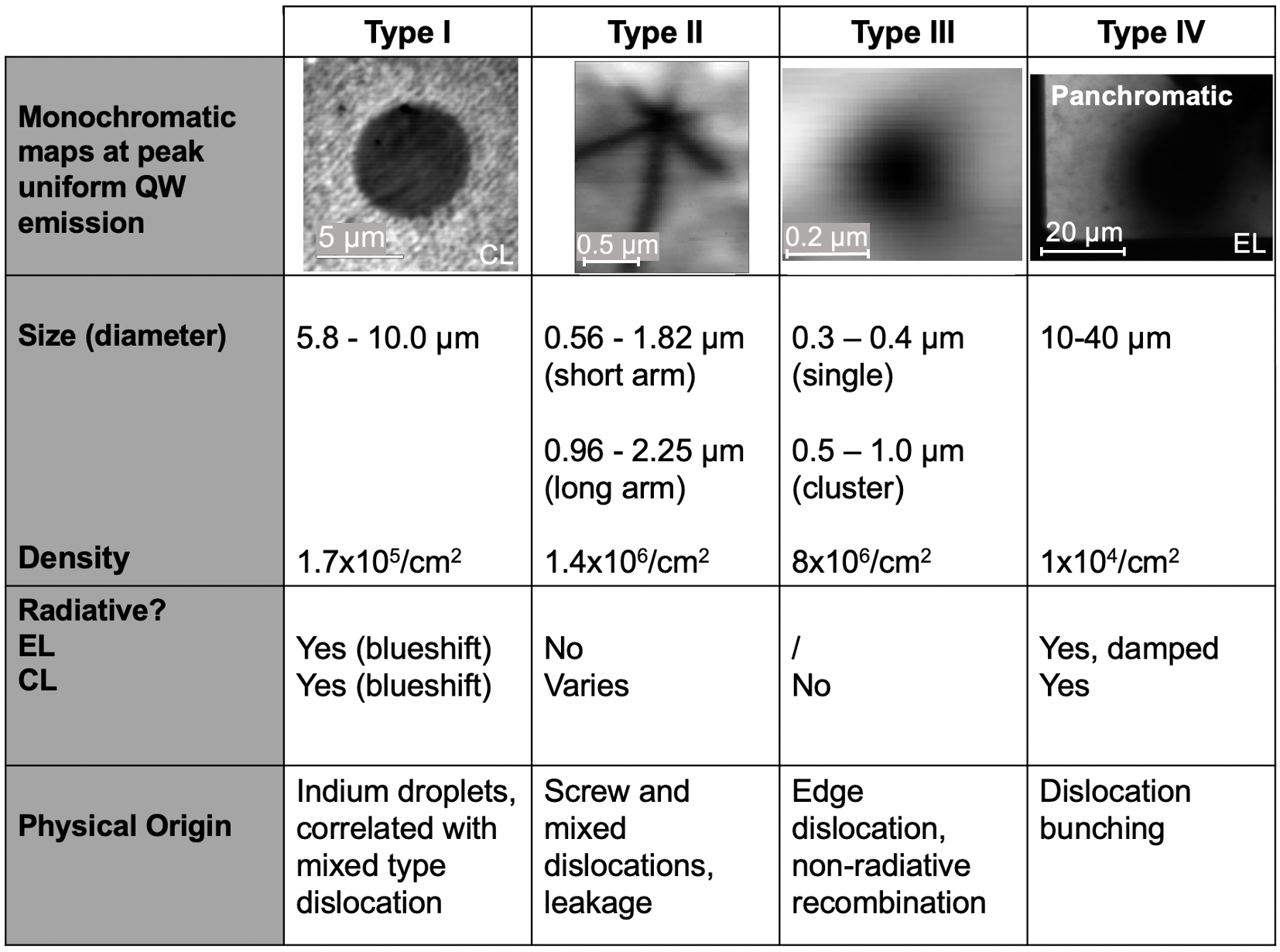}
\label{Table:1}}
\end{table}
%====================================================================

%====================================================================
\begin{figure*}[t]
\includegraphics[width=18cm]{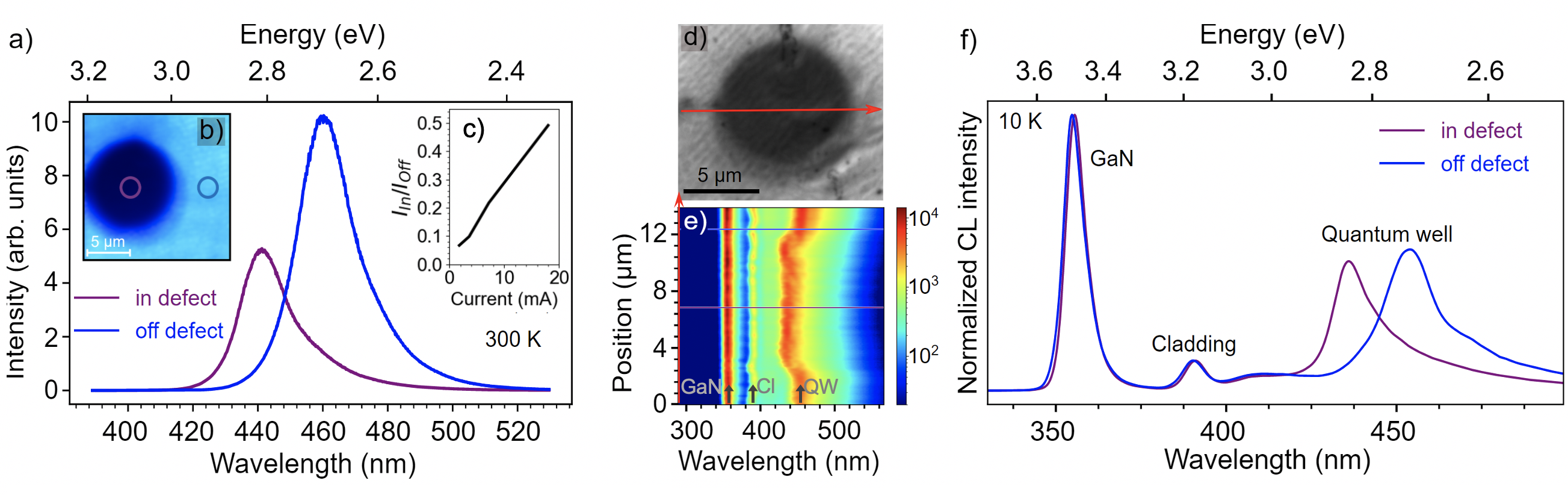}
\caption{(a) Localized electroluminescence spectra of a type I inhomogeneity and uniform region of the LED at room temperature and normalized injection current density of $J\approx 50$~A/cm$^{2}$. (b) Optical microscope image of the type I inhomogeneity, with the circles indicating the photon collection area.  (c) Ratio of emission intensity in and off the defect as a function of current. (d) Monochromatic CL image of a type I inhomogeneity recorded at the quantum well peak emission wavelength of 459.7 nm, with the red arrow indicating the position of the hyperspectral linescan at 10~K, which is shown in (e). In (e), the three arrows indicate the GaN, Cladding (Cl) and quantum well (QW) peaks. (f) Representative cathodoluminescence spectra off and on the type I defect recorded with an acceleration voltage of 7~kV from the positions marked by horizontal lines in (e). \label{fig:Figure_2}}
\end{figure*}
%====================================================================

The measurements reported here are performed on blue InGaN LEDs grown by plasma-assisted molecular beam epitaxy (PA-MBE) on c-plane Ge-doped bulk GaN substrates with dislocation density $5\times10^{5}-3\times10^{6}$~cm$^{-2}$. The specific growth details and characterization are described in \cite{lee_light-emitting_2020}. The two LEDs are buried tunnel homojunction (n-p-i-n) and AlN-interlayered tunnel junction  (n-i-p-i-n) structures which have identical active regions. For improved hole injection, LEDs using this design rely on interband Zener tunneling \cite{grundmann_multi-color_2007}. The latter also exploits polarization-induced band realignment at the tunnel heterojunctions to enhance the tunneling current \cite{simon_polarization-induced_2009}. Moreover, with low-resistance n-contacts both on the top and bottom electrodes, the design of these LEDs shows enhanced current spreading compared to a p-i-n LED, making it possible to do spatially resolved electroluminescence measurements. Both of the LEDs show similar inhomogeneity dimensions and density and most of the measurements reported here are done on the AlN-interlayered LED. Figure~\ref{fig:Figure_1}(a)-(b) shows the quantum heterostructure of the AlN-interlayered LED and the energy band diagram of the p-i-n segment at 3.0~V forward bias simulated by SiLENSe. All grown n-type GaN layers are Si-doped with a concentration of [Si]=10$^{19}$~cm$^{-3}$, where the p-GaN is Mg-doped with a concentration of [Mg]=$2\times10^{19}$~cm$^{-3}$. From bottom upwards, the unintentionally doped (UID) active region consists of a 22~nm In$_{0.07}$Ga$_{0.93}$N cladding layer, followed by three periods of 2.8~nm In$_{0.15}$Ga$_{85}$N quantum wells/7~nm In$_{0.07}$Ga$_{0,83}$N barriers, with an 18~nm In$_{0.07}$Ga$_{0.93}$N cladding layer on top. The diode current-voltage characteristics shown in figure \ref{fig:Figure_1}(c) indicate a forward turn on above $\sim$2.5 V and successful implementation of the buried tunnel junction.  The resulting MQW electroluminsecence peak emission is at $\approx 460$~nm or $h \nu \approx 2.7$~eV at room temperature as seen in Figure \ref{fig:Figure_1}(d) for various injection currents. Figure~\ref{fig:Figure_1}(e) shows optical microscope images of a 500$\times$100 $\mu$m$^{2}$ device with a first type of inhomogeneity clearly visible at 20~A/cm$^{2}$ as spots. The emission becomes increasingly uniform at higher current densities of 50~A/cm$^{2}$, which is further evidenced by Figure~\ref{fig:Figure_2}(c) which shows that these inhomogeneities brighten with respect of the rest of the (homogeneous) region of the LED as a function of current.

Spatially resolved electroluminescence measurements were performed at room temperature with a silicon photodiode detector. The collection area diameter, $D_{c}$, is limited by a $\times 20$ optical magnification and silica optical fiber core diameter of 50~$\mu$m, resulting in $D_{c} = 2.5~\mu$m. Monochromatic and spectrally-resolved cathodoluminescence maps were acquired with a Gatan MonoCL4 system using a high-sensitivity photomultiplier and a charge-coupled device as detectors. Using a 300 lines/mm grating and slits of 0.5~mm width, the spectral resolution can be estimated to be 5~nm. The system is mounted to a Zeiss Ultra55 scanning electron microscope operated at an acceleration voltage of 5--7~kV and equipped with a liquid-He cold-stage for measurements at cryogenic temperatures. The spectral maps were analyzed using the python package HyperSpy \cite{pena_hyperspyhyperspy_2020}. The spatial resolution is limited by the scattering of incident electrons and diffusion of excited carriers, which amounts to a few tens of nm.

We first analyse the largest inhomogeneity, which we call type~I, as shown in the optical microscope image of the LED under electrical injection in Figure \ref{fig:Figure_2}(b) and the monochromatic cathodoluminescence maps at the quantum well emission wavelength in Figure \ref{fig:Figure_2}(d) and Table \ref{Table:1}. They manifest as circular spots ranging from $6 \leq D_{I} \leq 10~\mu$m in diameter at a density of $N_{I} \approx 1.7 \times 10^{5}$~cm$^{-2}$. Local electroluminescence spectra taken at 300K are shown in Figure \ref{fig:Figure_2}(a) with corresponding collection area in and off the defect indicated by the circles in \ref{fig:Figure_2}(b), as well as the hyperspectral linescan by cathodoluminescence shown in \ref{fig:Figure_2}(e) with representative spectra in and off the defect shown in \ref{fig:Figure_2}(f). Clearly, The quantum well emission is blueshifted within the defect by $\Delta E \approx 120$~meV ($\approx 460$~nm to $\approx 440$~nm) relative to the spatially integrated peak at room temperature. We postulate that the blueshift is due to a perturbation of the active region that results from a decreased indium incorporation and reduced quantum well thickness in the region of the type I inhomogeneity. First, we discuss the perturbation due to a compositional variation. It is well known that the bandgap for the ternary In$_{\mathrm{x}}$Ga$_{1-\mathrm{x}}$N can be accurately described by a modified Vegard's law that exhibits a linear as well as quadratic dependence on the indium incorporation $\mathrm{x}$. Specifically,  the bandgap energy for In$_{\mathrm{x}}$Ga$_{1-\mathrm{x}}$N follows the equation:
\begin{equation}
    E_\mathrm{InGaN}(\mathrm{x}) = \mathrm{x} E_\mathrm{InN} + (1-\mathrm{x}) E_\mathrm{GaN} - b
    \mathrm{x} (1-\mathrm{x}),
\label{eqn_InGaN_Bandgap_Bowing}
\end{equation}
where $E_\mathrm{InN}$ is the bandgap of InN (0.65 eV at room temperature) and $E_\mathrm{GaN}$ is the bandgap of GaN (3.4 eV at room temperature), and $b$ is the bowing parameter for wurtzite InGaN. It has been shown that the value of b strongly depends on whether the InGaN grown on GaN is strained (b $\approx 1.3$~eV) or relaxed (b $\approx 2.9$~eV) \cite{orsal_bandgap_2014}. The perturbation to the quantum well transition energy in the inhomogeneous region with reduced indium incorporation  $\Delta \mathrm{x}$ compared to the rest of the (homogeneous) LED, is approximately equal to:

\begin{equation}
    \Delta E (\Delta \mathrm{x}) \approx \Delta \mathrm{x} (E_\mathrm{InN}-E_\mathrm{GaN}-b),
\label{eqn_Indium_Fluctuation_Perturbation}
\end{equation}

where the terms depending quadratically on the composition are neglected for low indium incorporated quantum wells.
On the other hand, the conduction and valence band bound energy eigenvalues in the quantum well can be approximated by the solutions to finite potential wells. The exact solution to Schr\"odinger equation of the finite well ground state energy follows the relation:

\begin{equation}
    \sqrt{\frac{E^{0}}{|E_{B}-E_{W}|}} = \cos{\frac{L \sqrt{2 m^{\ast} E^{0}}}{\hbar}},
\label{eqn_FinitWellSolution}
\end{equation}

where $E^{0}$ is the ground state energy of the conduction band or valence band electron, $E_{B}$ and $E_{W}$ are the barrier and well conduction band minimum or valence band maximum, $m^{\ast}$ is the electron or hole effective mass, and $L$ is the quantum well thickness. The energy band offset is larger for the conduction band than for the valence band and we make the approximation that the ratio between those offsets is 7:3. This implicit equation for $E^{0}$ can be separated by making an approximation in which the cosine term on the right side of Equation \ref{eqn_FinitWellSolution} is approximated by a polynomial \cite{de_alcantara_bonfim_exact_2005}:

\begin{equation}
  \cos x \approx f_{s}(x) \equiv \frac{1-(2 x / \pi)^{2}}{1+\frac{x^{2}}{2}\left(1-8 / \pi^{2}\right)} \quad(0 \leqslant x \leqslant \pi / 2).
\label{eqn_CosineApproximation}
\end{equation}

Solving for the electron and hole ground state eigenvalues $E^{0}_{e}$ and $E^{0}_{h}$ then yields:

\begin{equation}
  E^{0}_{e} + E^{0}_{h}  \approx \sum_{\substack{e, h}} \frac{\hbar^{2}}{2 m_{e,h}^{\ast} L^{2}}\frac{1+8 z_{e,h} / \pi^{2}-\sqrt{1+2 z_{e,h}}}{\left(32 z_{e,h} / \pi^{4} + 8 / \pi^{2} - 1\right)},
\label{eqn_Eigenvalue_QuantumWell}
\end{equation}
where $m_{e,h}^{\ast}$ are the electron and hole effective mass and $z_{e,h}$ is defined as $2\times \eta_{e,h}\times(L/\hbar)^{2}\times m_{e,h}^{\ast}\times|E_{B}-E_{W}|$ with $\eta_{e,h} \approx 7/10$ and $3/10$ for the conduction band and valence band, respectively.  

The combined effect of both decreasing the indium incorporation as well as shrinking the quantum well thicknesses increases the emission energy within the type I defect compared to the rest of the LED, as seen in Figure \ref{fig:Figure_2}(a).  Given equations \ref{eqn_Indium_Fluctuation_Perturbation} and \ref{eqn_Eigenvalue_QuantumWell}, we approximate the 120~meV energy shift due the effect of reduced indium incorporation or reduced quantum well thickness. The long emission wavelength of 460~nm for the $\mathrm{x} \approx 0.15$ quantum wells suggests the InGaN is relaxed. By equation \ref{eqn_Indium_Fluctuation_Perturbation} and using $b \approx 2.9$~eV, the sole effect of a reduction of indium incorporation leads to $\Delta \mathrm{x} \approx -0.021$, reducing the quantum well indium content from $\approx 15$\% to $\approx 12.9$\%. On the other hand, by equation \ref{eqn_Eigenvalue_QuantumWell}, using $m_{e}^{\ast} \approx 0.2 m_{e}$ and $m_{h}^{\ast} \approx 0.8 m_{e}$ where $m_{e}$ is the free electron mass, the 120~meV blueshift by thinning of the quantum wells alone is calculated. This would result in a significant reduced well thickness of $\Delta L \approx 1.50$~nm in the type I defect, resulting in $L \approx 1.3$~nm. Hence, $-0.021 \leq \Delta \mathrm{x} \leq 0$ and $0 \leq \Delta L \leq 1.50$~nm. 

This is in fair congruence with LED heterostructure simulations performed using SiLENSe, where the quantum confined Stark effect (QCSE) arising from polarization fields is taken into account as well. The energy-shift then corresponds to $-0.029 \leq \Delta \mathrm{x} \leq 0$ and $0 \leq \Delta L \leq 1.53$~nm.
It must be noted that this type of inhomogeneity differs in nature from nanoscopic emission inhomogeneities which are in part responsible for spectral broadening, as well as microscopic emission inhomogeneities on the  smaller order of $\leq 10$~meV. Both can also be explained by compositional fluctuations as well as quantum well monolayer thickness variations. The nanoscopic fluctuations have been held responsible for the high efficiency achieved in quantum well based InGaN blue LEDs, whereas the microscopic fluctuations are in general not desirable \cite{yang_influence_2014-1,sakaki_impact_2019}.

%====================================================================

\begin{figure}[t]
\includegraphics[width=8.7cm]{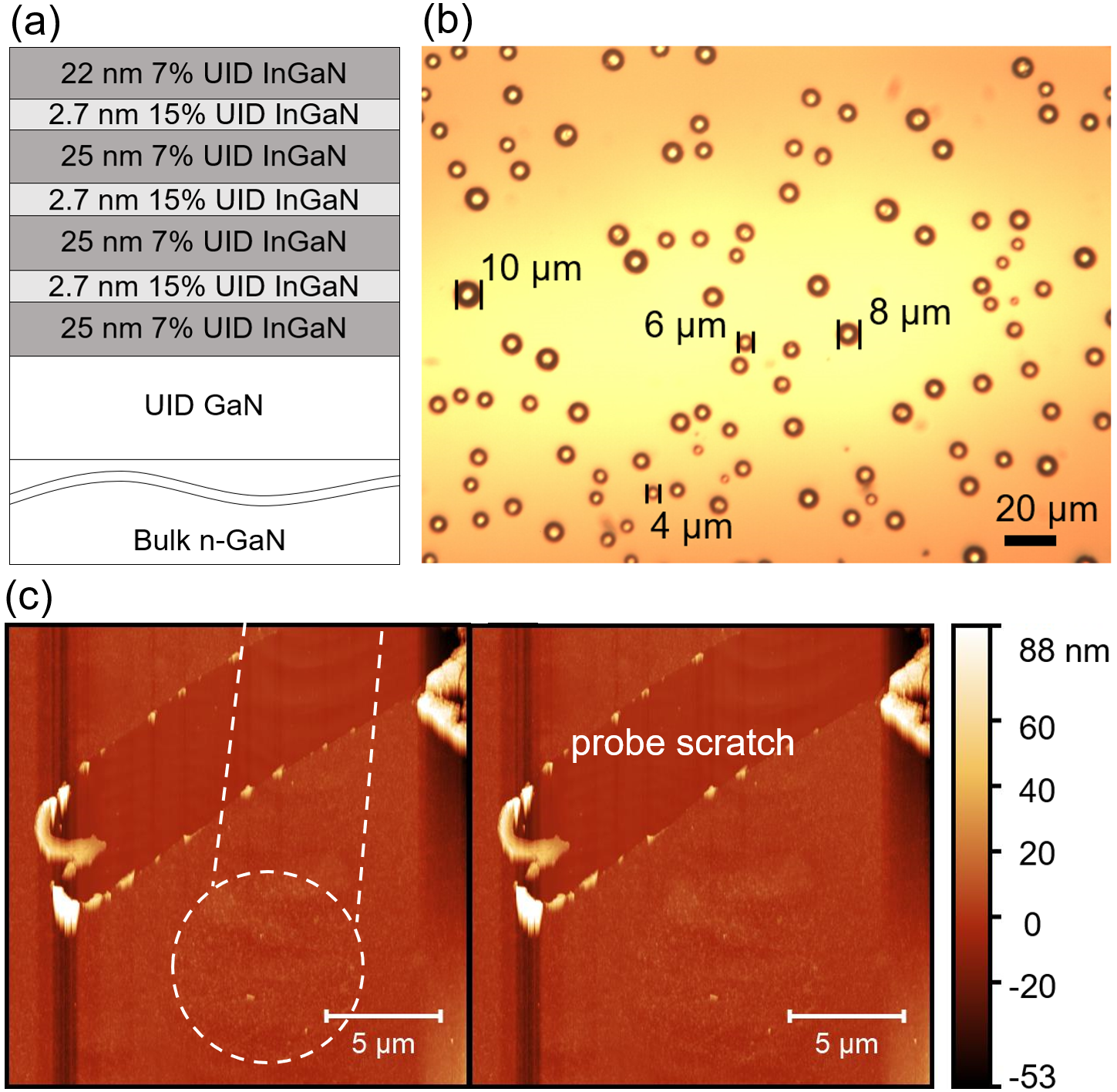}
\caption{(a) Sketch of an independently grown multiple quantum well (MQW) structure, with InGaN layers identical to the active region of the LED in Figure \ref{fig:Figure_1}(a), but with no GaN layer on top. (b) Optical microscope image of the surface of the MQW structure in (a), where the indium droplets were deliberately not desorbed. (c) Atomic force microscopy (AFM) image, taken at the location of a type~I inhomogeneity on the LED surface. The surface morphology is wavy leading towards and a depression is located at the type~I inhomogeneity indicated by the dashed circle. The rough slanted line is a scratch caused by a probe during electroluminescence measurements, and it also serves as a marker to identify the relative position of the inhomogeneity.  \label{fig:Figure_3}} 
\end{figure}

The type I inhomogeneity is postulated to originate from altered InGaN growth conditions at the location of indium droplets. Such droplets are known to form during PA-MBE under metal (indium) rich growth conditions. Zheng et al. have identified the formation and movement of In droplets on the c-plane growth surface \cite{zheng_effect_2018}. The reported size of the droplets coincide with the size of the type I inhomogeneities discussed here. To confirm this observation, an independent MQW sample was grown with the same indium incorporation in its cladding, quantum well and barrier layers as the LED structure, but with no GaN layer on top, and where the indium droplets were deliberately not desorbed, as indicated in Figure \ref{fig:Figure_3}(a). Indeed, Figure \ref{fig:Figure_3}(b) shows that the density and size of the type I inhomogeneities are in congruence with the density and size of indium droplets observed on the surface of the sample in Figure \ref{fig:Figure_3}~(a). Furthermore, AFM scans of the surface of the processed LED shown in Figure \ref{fig:Figure_1}(a) in the region of the type I inhomogeneity show a circular depression. The surface morphology is often wavy leading towards and into the type I inhomogeneity, which is highlighted by the dashed circle in figure \ref{fig:Figure_3}(c). This is likely indicative of the indium droplet migrating during growth of the InGaN layers. The wavy nature is due to the 150 nm n-GaN contact layer grown on top of the top InGaN cladding layer. The rough slanted line is a scratch caused by a probe during electroluminescence measurements, and it also serves as a marker to identify the relative position of the inhomogeneity. The observed morphology, along with the blueshift of the type I inhomogeneity then suggest that the incorporation of In and the growth rate are indeed decreased in the region of an indium droplet during the growth of the InGaN active region. 

%====================================================================
%====================================================================

A model for the variation of the indium composition and growth rate is now discussed.  It is well known that the indium incorporation x in PAMBE-grown InGaN follows the relation

\begin{equation}
    \mathrm{x} = 1 - \Phi_\mathrm{Ga}/\Phi_\mathrm{N}^{inc},
\label{eqn_growth_fluxes}
\end{equation}

where $\Phi_{\mathrm{Ga}}$ and $\Phi_{\mathrm{N}}^{inc}$ are the incident gallium and incorporated nitrogen fluxes respectively \cite{monroy_incorporation_2003,siekacz_growth_2008}. The nitrogen that actually incorporates into InGaN, $\Phi_{\mathrm{N}}^{\mathrm{inc}}$, is approximately equal to the supplied nitrogen flux $\Phi_{\mathrm{N}}^{\searrow}$ only for low indium compositions ($\lesssim$~10\%). This is because the decomposition of InGaN depends exponentially on an activation barrier, which is larger for GaN (3.6 eV) than InN (1.92 eV) \cite{averbeck_quantitative_1999,grandjean_gan_1999,gallinat_role_2009}. Furthermore, the gallium adlayer present during 2D growth does not desorb at the 660~$^{\circ}$C temperature used for metal rich growth for the active region of the LEDs \cite{he_gallium_2006}. Hence, all of the incident Ga in Equation \ref{eqn_growth_fluxes} is incorporated. For high $\Phi_{\mathrm{Ga}}$/$\Phi_{\mathrm{N}}^{\searrow}$ ($\gtrsim$~0.9), corresponding to low indium composition InGaN growth, this then means that the incident nitrogen flux can be approximated to be completely incorporated into InGaN. However, this is not true for lower $\Phi_{\mathrm{Ga}}$/$\Phi_{\mathrm{N}}^{\searrow}$ where there will be a nitrogen excess. 

Indeed, Turski et al. have shown that at low gallium fluxes, initially formed InN bonds are susceptible to decomposition. The released nitrogen can either become part of the incident nitrogen flux and form another InN or GaN bond with adjacent indium or gallium adatoms, or evaporate and be lost indefinitely \cite{turski_nonequivalent_2013}. Hence, there will be a nitrogen excess such that $\Phi_{\mathrm{N}}^{\mathrm{inc}} < \Phi_{\mathrm{N}}^{\searrow}$. Finally, it is important to note that for the metal rich growth condition ($\Phi_{\mathrm{Ga}} + \Phi_{\mathrm{In}} > \Phi_{\mathrm{N}}^{\mathrm{inc}}$) for a given growth temperature $T_{\mathrm{G}}$, the growth rate is determined by $\Phi_{\mathrm{N}}^{\mathrm{inc}}$. We employ this as given knowledge to explain the growth of the InGaN MQWs underneath the indium droplet and the origin of the blueshift. 

First, it should be a justifiable assumption that the gallium flux is not reduced underneath the indium droplet since the diffusion length of gallium adatoms is large in the 2D growth mode \cite{skierbiszewski_growth_2004}. The behavior of the measured cladding (Cl) peak by cathodoluminescence at $\approx 390$~nm or 3.2~eV, as indicated by middle the black arrow in the hyperspectral linescan in Figure~\ref{fig:Figure_2}(e), provides an important clue to understand the growth dynamics underneath the droplet. First and foremost, this peak is not visible in the electroluminescence spectrum at any measured injection current density because of the enhanced injection efficiency and negligible carrier overflow for the inverted LED structure \cite{bharadwaj_enhanced_2020, van_Deurzen_Enhanced}. However, from the low temperature cathodoluminescence data, the $\approx390$~nm peak from the In$_{0.07}$Ga$_{0.93}$N cladding layers are visible, but their emission energy is not blueshifted compared to the homogeneous region of the LED. The two InGaN peaks are labeled in Figure \ref{fig:Figure_2}(c) by the grey arrows. This discrepancy in energy-shift between the two InGaN peaks suggests that the growth environment underneath the indium droplet is only changed for the MQW which have a higher indium composition In$_{0.15}$Ga$_{0.85}$ and not in the barriers and cladding layers. Specifically, this implies that the reincorporation of nitrogen atoms stemming from decomposed InN bonds is reduced underneath the droplet compared to the rest of the LED, resulting in a reduced $\Phi_{\mathrm{N}}^{\mathrm{inc}}$ and increased $\Phi_{\mathrm{Ga}}$/$\Phi_{\mathrm{N}}^{\mathrm{inc}}$, explaining the reduced growth rate and indium incorporation of the quantum wells underneath the droplet. 

\begin{figure}[t]
\includegraphics[width=8.7cm]{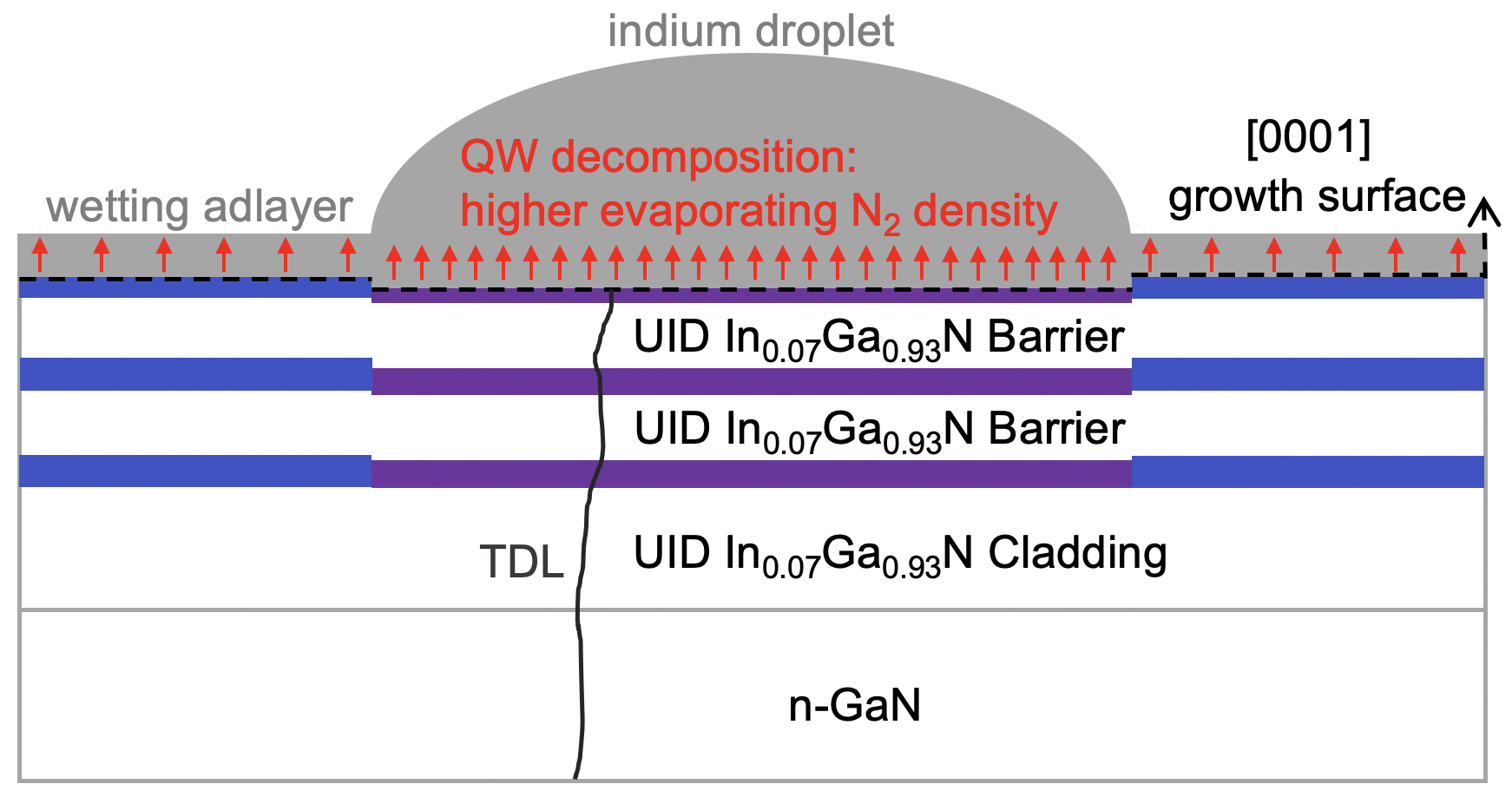}
\caption{Illustration of MQW growth and the altered growth conditions underneath an indium droplet.  \label{fig:Figure_4}}
\end{figure}

\begin{figure}[t]
\includegraphics[width=8.7cm]{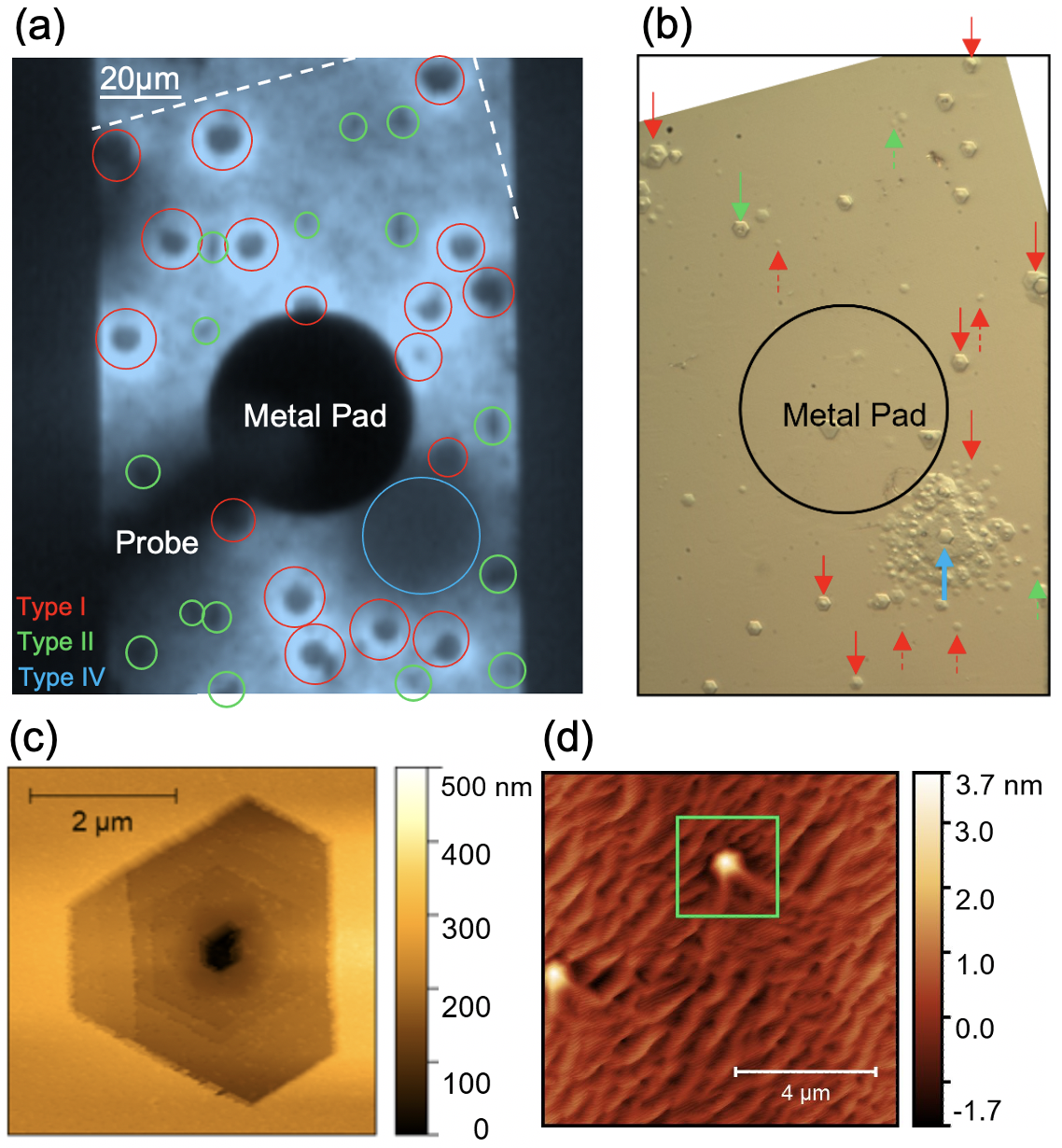}
\caption{Correlation between dislocations and emission inhomogeneities. (a) Optical microscope image of the LED in its electroluminescent state, with type I (red), discernible (larger) type II (green), and type IV (blue) inhomogeneities indicated with circles. (b) Optical microscope image of the same region, after KOH etching at 200~$^{\circ}$C for 10 minutes. Arrows indicate the location of chemical etch pits revealed at the location of the type I, II and IV inhomogeneities. Most of the type I inhomogeneities are pinned at the location of mixed type dislocations.  For type IV (blue circle), dislocation bunching is observed and the etch rate is increased in this region (blue arrow), revealing a cluster of pits. Merely 3 of the discernible type~II inhomogeneities show up as etch pits. The smaller type II and III inhomogeneities cannot be resolved from the OM image, and its correlation with screw and edge dislocations is discussed separately. (c) AFM image of a chemical etch pit corresponding to a mixed type dislocation, which was revealed at the location of a type I inhomogeneity. (d) AFM image at the location of a type~II inhomogeneity, indicated by the green square, after the LED growth and before processing or KOH etching. The V-shaped and spiral hillock are indicative of a screw dislocation. \label{fig:Figure_5}}
\end{figure}

As a potential explanation, the surface underneath the droplets becomes increasingly nitrogen rich and enhances the formation of molecular $N_{2}$, which, unlike the atomic or ionic Nitrogen, is not an active growth species. This reduces the reincorporation of N with In to form InN. The altered active region and growth conditions due to the indium droplets are illustrated in Figure~\ref{fig:Figure_4}.
%====================================================================
%====================================================================
From the density and size of the type I inhomogeneities, they are calculated to cover $\approx13$~\% of the LED area. This is significant enough to undesirably broaden the spatially-integrated spectrum of the LED, especially at higher normalized current densities, as can be seen from the shoulder in the log-scale spatially-integrated emission spectrum in Figure~\ref{fig:Figure_1}(d). Furthermore, the type I region is transparent to the modes for stimulated emission in a Laser Diode, reducing its gain.  The right inset of Figure~\ref{fig:Figure_2}(c), which shows the ratio of the emission intensity on and off the defect, suggests that the droop in the homogeneous region occurs earlier than the droop in the type I region as a function of the normalized current density. This is due to the fact that the current density in the type I region is lower for any given total current as it has a larger bandgap, essentially forming a 'quantum hill'. The density of droplet-related defects can be minimized by using an optimal indium flux during growth. On the other hand, due to growth temperature fluctuations or instability, complete removal of this defect for indium-rich growth is challenging.
%====================================================================
Another interesting observation about the type I inhomogeneities is the correlation of their location with dislocations that are revealed by hot potassium hydroxide (KOH) etching, as shown by the optical microscope image of the LED in its in Figure \ref{fig:Figure_5}(a) and (b). The KOH etch was performed at 200~$^{\circ}$C for 10 minutes \cite{xu_acid_2002}. Specifically, most of the type I inhomogeneities that correlate to chemical etch pits correspond to mixed type dislocations such as the one imaged by AFM in Figure \ref{fig:Figure_5}(c). Zheng et al. have reported that the location of indium droplets are correlated with dislocation density for MBE growth, again suggesting that the type I defects are due to indium droplets \cite{zheng_effect_2018}. It is uncertain whether the droplets form at the location of the dislocation or form elsewhere and migrate along the growth surface to eventually be pinned at the dislocation due to strain fields, or both. The latter mechanism can explain the observed wavy morphology shown in Figure \ref{fig:Figure_3}(c). The correlation suggests that the surface energy of the droplets is reduced at the location of the dislocation due to the strain fields of those dislocations. From another LED grown on a bulk GaN substrate with a lower specified dislocation density of $5\times10^{4}$~cm$^{-2}$, the type I density is reduced to $N_\mathrm{I}$ = $3\times10^{4}$~cm$^{-2}$. This further corroborates that the indium droplet density positively correlates with the dislocation density of the substrate.

%====================================================================
%====================================================================

%====================================================================
\begin{figure*}[t]
\includegraphics[width=18cm]{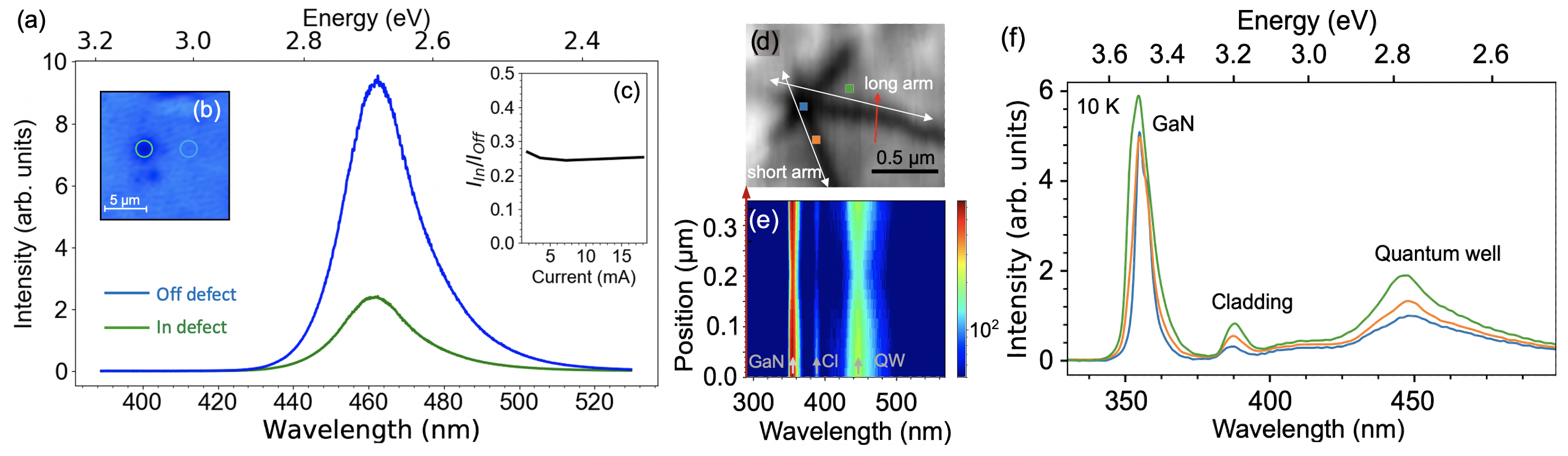}
\caption{(a) Localized electroluminescence spectra of a type II inhomogeneity at room temperature and a normalized current density of 50~A/cm$^{2}$, with (b) the 2.5~$\mu$m diameter circles indicating the photon collection area. (c) Ratio of the defect output intensity to that of the rest (homogeneous region) of the LED. (d) Integrated cathodoluminescence intensity map. The defect is star-shaped and is in general not C6 symmetric, and its shape is defined by two lines: the short and long arm as indicated by white arrows. (e) Hyperspectral linescan through an arm of the defect, extracted along the red arrow in (d). The GaN, cladding (Cl) and QW peaks are indicated by the grey arrows. (f) Selected spectra from the CL map of the defect recorded at an acceleration voltage of 5~kV, at different positions on and off the arms and at the center. \label{fig:Figure_6}}
\end{figure*}
%====================================================================

Next, we discuss another type of inhomogeneity, which we refer to as type II. They manifest as smaller (than type I) dark spots of diameter $D_\mathrm{II} < 5$~$\mu$m under current injection.  The discernable (larger) type II dark spots are indicated by green circles in Figure~\ref{fig:Figure_5}. From cathodoluminescence, the density of type II spots is measured to be $N_\mathrm{II} \approx 1.4 \times 10^{6}$~cm$^{-2}$. AFM scans in the region of the type~II defects prior to KOH etching indicate that the surface morphology is that of a spiral with star-shaped extensions, as shown by the green square in \ref{fig:Figure_5}(d). This morphology is known to correspond to screw dislocations for MBE-grown GaN \cite{qian_opencore_1995, heying_dislocation_1999}. However, after etching, merely a small fraction of the screw dislocations show up as V-pits in Figure \ref{fig:Figure_5}(b), meaning that the KOH etch at 200~$^{\circ}$C for 10 minutes is not optimized to reveal V-pits corresponding to screw dislocations. 
%====================================================================
An interesting observation is that these screw dislocations are dark in the electroluminescence maps but not always in the cathodoluminescence maps at the quantum well emission energy. Figure \ref{fig:Figure_6}(a)-(b) shows the quantum well spectrum in and off the defect under electrical injection at 50~A/cm$^{2}$. The intensity is damped within and around the defect and the ratio of output intensity relative to the rest of the LED does not change as a function of current, as shown in Figure \ref{fig:Figure_6}(c). This suggests that the center of the type II defect is truly dark in EL and the collected spectrum at these positions is background from the rest of the LED since the size of the collection area is similar to the size of the type I spot. On the other hand, the type II defects display a hexagonal star-like appearance in maps of the QW emission taken by cathodoluminescence, as shown in Figure \ref{fig:Figure_6}(d) and Table \ref{Table:1}.  Figures \ref{fig:Figure_6}(d)--(f) show an integrated intensity map, hyperspectral linescan and selected emission spectra of a dark type II inhomogeneity respectively, taken by cathodoluminescence. It is noticed that in general, the type II is not C6 symmetric and we can define its shape by two axes, which we call the short and long arm as indicated in Figure~\ref{fig:Figure_6}(d) by the white double arrows. The arms can clearly be attributed to the ridges of the hillocks as can be seen from the AFM scan in Figure~\ref{fig:Figure_5}(d), showing the surface morphology of two type~II defects. From this figure, it is clear that the asymmetry is due to atomic steps and a local miscut of the sample surface. The measured short and long arms range from 0.56 $\mu$m to 1.82 $\mu$m and 0.96 $\mu$m to 2.25 $\mu$m, respectively. This ridge formation on the c-plane surface and its dependence of the miscut angle and relation to dislocations has been discussed in GaN grown by metal organic vapor phase epitaxy (MOVPE) \cite{oehler_surface_2013}. Figures \ref{fig:Figure_6}(e) and (f) show that the intensity of the InGaN peaks within the arms are damped compared to off the arms. 
%====================================================================
Another interesting observation is that the intensity of the GaN peak is not damped as strongly as the InGaN peaks and a that there is a slight energy-shift of the InGaN peaks at the center of the defect and within the arms. For the center and the arms, indium segregation could happen due to growth on the hillock facets. At the center, the energy shift of the InGaN peaks could also be explained by indium segregation due to the strain fields induced by the screw dislocation \cite{lei_role_2010, horton_segregation_2015}. Some of the type II defects do not exhibit damped InGaN peaks at the arms and center, yet others exhibit only a dark center in the cathodoluminescent state. The dark center could indicate that they are in fact mixed type dislocations where the dark center is due to non-radiative recombination as described in the following paragraph. Other potential explanations for the dark center are that they are a result of different radiative properties depending on the core structure of the screw dislocation or the presence of voids.

%====================================================================
\begin{figure*}[t]
\includegraphics[width=16cm]{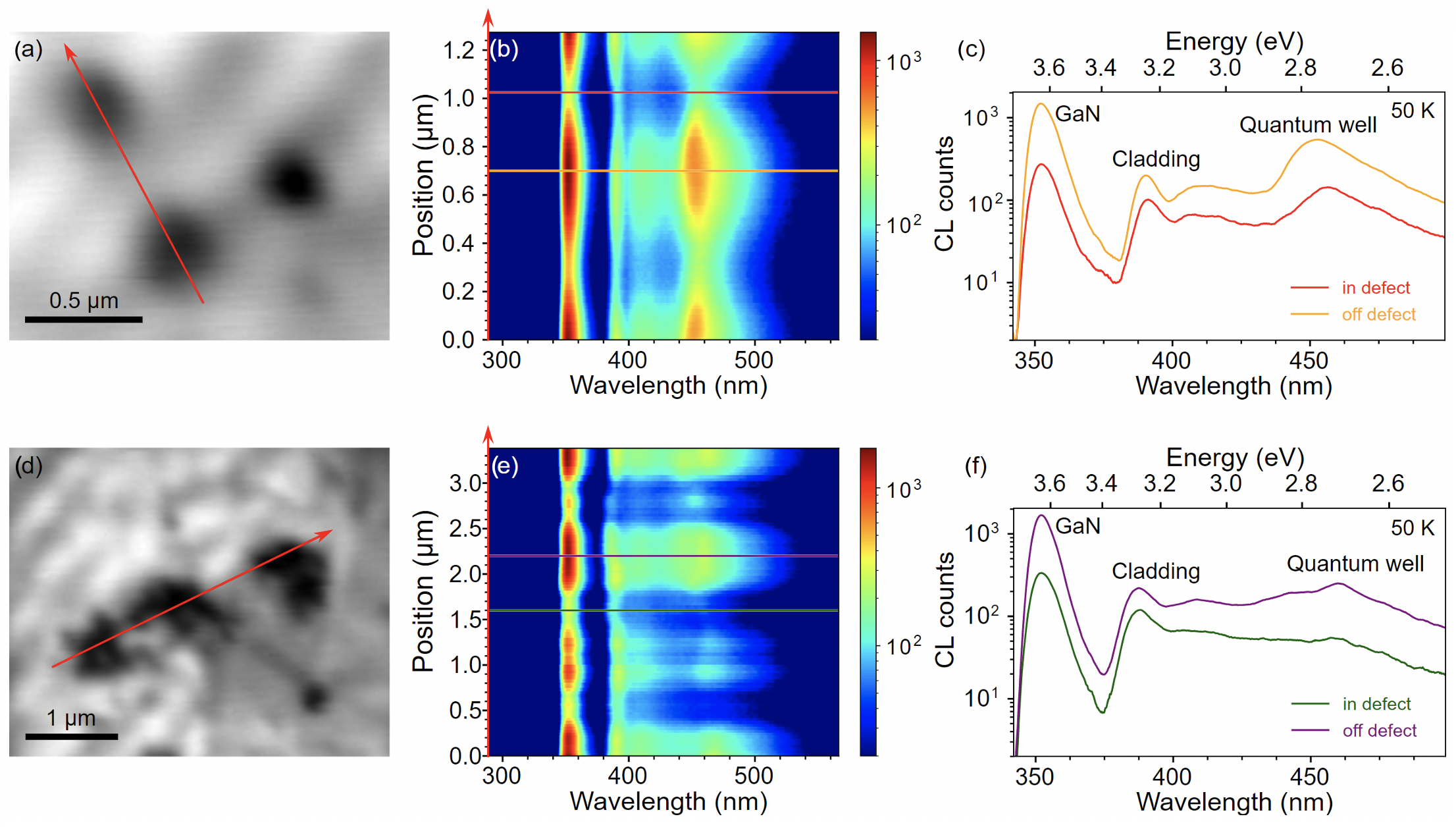}
\caption{Cathodoluminescence measurements for the small type~III defects recorded with an acceleration voltage of 5~kV. (a) Integrated cathodoluminescence intensity map. (b) Hyperspectral linescan extracted along the red arrow in (a). (c) Selected spectra from the linescan for the positions indicated in (b).  Similarly, (d)--(f) show the integrated intensity map, hyperspectral linescan and selected spectra for clusters of type III defects. \label{fig:Figure_7}}
\end{figure*}
%====================================================================

The difference seen in the radiation properties from electroluminescence and cathodoluminesncence is a result of the distinct nature of the transport of electrons and holes in the two methods of excitation. In electroluminescence, the electrons and holes are injected by drift and diffusion into the quantum wells from the opposite ends of the active region through the doped cladding layers and generated by thermal excitation from dopants. In cathodoluminescence, the electron-hole pairs are generated locally by the inverse photoelectric effect and diffusion towards the active region occurs in the same direction, allowing for electrons and holes to be injected into the active region. Hence, the reason for suppressed emission in the electroluminescent state stems from an alteration of the electronic properties of the doped GaN cladding layers, limiting the transport of holes and/or electrons to the active region by electrical injection alone. Specifically, Usamie et al. and Nakano et al. have proven that Mg diffuses along threading screw dislocations (TSDs) and forms a Mg-TSD complex that turns p-type GaN into n-type \cite{usami_direct_2019,nakano_screw_2020}. Depending on the diffusion length of magnesium along the threading screw dislocation into the unintentionally doped InGaN active region and the electronic properties of the Mg-screw complex in InGaN, the p-i-n LED will be converted to n-i-n or n-n-n in the vicinity of the threading screw dislocation, limiting the amount of holes in the active region for radiative recombination and (in the n-n-n case) forming leakage paths.  
%====================================================================

We next discuss the more prevalent type~III inhomogeneity that manifests as small dark regions throughout the LED cladding layers and active region with a diameter $0.3 \leq D_\mathrm{III} \leq 0.4$~$\mu$m and a density of $N_\mathrm{III} = 8 \times 10^{6}$~cm$^{-2}$. The measurements in Figure \ref{fig:Figure_7}(a)--(c) show an integrated intensity map, hyperspectral linescan and selected spectra for an area of the LED consisting of three isolated type~III defects taken by cathodoluminescence. Within the type~III region, all peaks in the spectrum of the LED are quenched. The type III inhomogeneities are too small to be resolved by the localized electroluminescence setup used in this work. Since the density $N_\mathrm{III}$ is a factor of $\approx 6$ larger than the type II screw dislocation density $N_\mathrm{II}$, we attribute the type~III to edge dislocations, because the elastic energy of edge dislocations is lower than for screw dislocations \cite{elsner_theory_1997}. The reason for the suppressed emission around the threading edge dislocation can then be explained by the formation of non-radiative recombination centers at the threading dislocation in both GaN and InGaN \cite{yao_correlation_2020, bojarska-cieslinska_role_2021, lahnemann_carrier_2020}. Regions in which multiple type III defects bunch together and form clusters are also observed. For these defects, similar quenching of the peaks is observed, yet in a larger radius around the center of the defect. This is evidenced by the cathodoluminescence measurements that are shown in Figure \ref{fig:Figure_7}(d)--(f). 
%====================================================================

Finally, we discuss a fourth type of inhomogeneity, type~IV. They come at a density of $N_\mathrm{IV} \approx 10^{4}$~cm$^{-2}$ and their diameters range from 10 $\mu$m $\leq D_\mathrm{IV} \leq 40$~$\mu$m. Panchromatic images of type~IV defects under current injection are shown by the blue circle in \ref{fig:Figure_5}(a) as well as in Table \ref{Table:1}. Similar to type~II, the intensity is damped in the electroluminescent state but not the cathodoluminescent state, suggesting that the InGaN active region is radiatively intact, but the electronic characteristics of the doped GaN cladding layers are altered in this region. Besides the much larger size, the differences between the type~IV with respect to type~II, are i) that, similar to type~I, the electroluminescence intensity increases with injection current and ii) its etch pit profile. Shown with the blue arrow in Figure \ref{fig:Figure_5}(b) is the KOH-etched surface of a type~IV defect, corresponding to the blue circle in \ref{fig:Figure_5}(a). Clearly, the etch rate is increased and a cluster of V-pits appear in the region of the type~IV, indicating dislocation bunching. Possible explanations for the observed radiative behavior could include the altered doping characteristics of Mg due to the Mg-TSD complex, reduced Si or Mg incorporation in the doped layers in this region, tunnel junction inhomogeneities or reduced mobility of mobile carriers and hence reduced current densities due to scattering by the high density of threading dislocations. As an alternative explanation, the type~IV might be due to surface damage resulting from processing the LED, resulting in the increased KOH etch rate.

In summary, as indicated in Table \ref{Table:1}, we have identified four types of dislocation-correlated emission inhomogeneities that manifest in InGaN LEDs grown by MBE: 

Type I are unique to MBE-grown LEDs and are indium droplet related alterations of the active region which result in a blueshift of the peak quantum well emission. Their location is correlated with mixed type dislocations. With a reduced indium incorporation and thinner quantum wells which results in a larger bandgap, the turn-on voltage is increased and current density is reduced in this region. 

Type II are screw dislocations for which the quantum well emission is dark in electroluminescence, but only slightly damped in cathodoluminescence for some defects, and is due to the altered electronic properties of the p-GaN. 

Type III are edge dislocations which are dark both in cathodoluminescence and electroluminescence, though with a much smaller radius, due to non-radiative recombination. 

Type IV exhibits a large region of suppressed electroluminescence radiation and is revealed to be a region of a high density of dislocations.

An unresolved question in III-Nitride semiconductor heterostructures used in photonic devices is the discrepancy between the internal quantum efficiencies (IQE) of MBE-grown compared to MOCVD-grown spontaneous emitters such as LEDs. This work indicates that the effect of the inhomogeneities reported here, and in particular dislocations, have a merely a slight effect on the overall IQE of the blue InGaN LED. Specifically, type~I broadens the J-IQE curve compared to a hypothetical LED without the inhomogeneity, where type II to IV are not unique to MBE-grown LEDs. From the measured intensity spectra and densities of the inhomogeneities, type II and III combined are calculated to reduce the IQE of the LED by $\approx$ 4$\%$ compared to a hypothetical LED with no dislocations. Hence, the discrepancy in IQE is likely related to point defects, impurities or an unidentified complex. Recent advancements in N-polar MBE growth have shown a substantial improvement in IQE of photonic structures by increasing the growth temperature and employing nitrogen-rich conditions \cite{turski_nitrogen-rich_2019}. This new, exotic approach re-enables the use of N-polar substrates for nitride emitters. Moreover, due to the lack of an indium wetting layer during growth, conducting such process is simpler and similar to MOVPE. Though the IQEs of MBE-grown nitride heterostructures have been low compared to those grown by MOCVD, certain advantages such as buried Mg-doped p-type GaN layers, and higher indium incorporation due to lower growth temperatures offer some unique capabilities in the future, if the low IQE problem can be overcome in a reproducible manner.
\\
%====================================================================
\null \quad The authors acknowledge support from the Cornell Center for Materials Research with funding from the NSF MRSEC program (No. DMR-1719875) and AFOSR Hybrid Materials MURI under Award No. FA9550-18-1- 0480. The authors further acknowledge the Cornell NanoScale Facility for device fabrication, supported by the National Science Foundation with grant No. NNCI-1542081 and RAISE-TAQS 1839196, as well as MRI 1631282.
%====================================================================

\pagebreak

\bibliography{main}

\end{document}